# Wind Turbine Gearbox Condition Based Monitoring

Alan Rezazadeh
Applied Research and Innovation Services
Southern Alberta Institute of Technology, 1301 – 16 Avenue NW
Calgary, AB, Canada T2M 0L4

Corresponding author Alan Rezazadeh (alan.rezazadeh@sait.ca)

Terms: CBM, Clustering, Normal Mixture, Power Curve,

## 1. Abstract

The main objective of this paper is finding effective gearbox condition monitoring methods by using continuously recorded monitoring SCADA (Supervisory Control and Data Accusation) data points. Typically for wind turbine gearbox condition monitoring; temperature readings, high frequency sounds and vibrations in addition to lubricant condition monitoring have been used. However, collection of such data, require shutting down equipment for installation of costly sensors and measuring lubricant quality. Meanwhile, operational data usually collected every 10 minutes, comprised of wind speed, power generated, pitch angle and similar performance parameters can be used for monitoring health of wind turbine components such as blades, gearbox and generator. This paper uses gear rotational speed for monitoring health of gearbox teeth; since gearbox teeth deterioration can be measured by monitoring rotor to generator rotation ratios over extended period of time.

As nature of wind is turbulent with rapid fluctuations, a wind turbine may operate in variety of modes within relatively short period of time. Monitoring rotational speed ratio over time, requires consistent operational conditions such as wind speed and torques within the gearbox. This paper also introduces the concept of clustering such as Normal Mixture algorithm for dividing operating datasets into consistent subgroups, which are used for long term monitoring.

## 2. Introduction

As global net zero emission targets are becoming more critical to meet within the specified timeframes by variety of environmental agencies, wind turbines are considered a viable alternative to traditional electricity generation technologies such as natural gas or coal electric turbines. However, wind turbines also present a number of challenges such as gearbox reliability, high nonproductive time, higher than expected maintenance cost or shorter than planned productive life expectancy (Pandit & Infield, 2018b).

Industrial gearbox performance and reliability has been investigated for decades, which provide in-depth understanding of potential causes of wear and fatigue resulting major failures (Z. Zhang et al., 2017, Flodin & Andersson, 1997). Wind turbine gearboxes, offer a unique set of challenges, such as difficult maintenance space in nacelle, transient loads, extreme accelerations/decelerations and load reversals (Y. Zhang et al., 2019). Installing sensors for traditional methods of condition monitoring requires extended shutdowns in addition to extra cost of expensive equipment, which adds to the cost of operating wind turbines (Duda et al., 2018). In general wind turbine industry has been facing increased costs and decreased productivity due to premature gearbox failures and high maintenance costs (Turnbull et al., 2020).

Wind turbines comprised of many complex components; while, gearboxes have one of the highest failure ratios among the components (Wilkinson et al., 2014). Gearbox is approximately 13% of overall capital cost of an onshore wind turbine, while incurs costly maintenance over time in addition to nonproductive time due to taking the wind turbines off line for repairs (Qiu et al., 2017). Therefore, condition monitoring is critical to maximize efficiency of gearboxes, applying repairs and maintenance only when needed based on quantitative measurements (Du et al., 2017).

Although wind turbines have been designed and manufactured to withstand a specified wind speed, temperature and working condition, often wind turbines fail meeting the life expectancy or may require extra maintenance than defined by manufacturers (Qiu et al., 2017). Equipment failures may include any component of a wind turbine; however, typical sources of failures are gearboxes, generators, and blades (Turnbull et al., 2020). Gearboxes are expensive and complex component of wind turbines, which may fail due to multiple reasons. Typical failures of gearboxes are bearings (70%), gears (26%) and other reasons (4%) (Kim et al., 2011). Some reasons for unexpected high failure rates of wind turbines could be due to relatively evolving industry, increasingly larger size of turbines and still relatively unknown nature of wind forces and turbulences (Wang et al., 2016).

The main objective of this study is finding a quantitative method, monitoring wind turbine gearbox status using rotor to generator speed ratio (Liu et al., 2020). Although gears are made from solid steel; however, the touch point between gears change as a result of wear which affects minute change to speed ratio of rotor to generator (Wang et al., 2016). This ratio is then used as a monitoring indicator of wear within the gear teeth which are physically difficult and expensive to be monitored, since the turbine needs to be shut down for the inspection (Wang et al., 2016).

The concept of monitoring speed ratio seems to be trivial due to its simplicity; however, in order to monitor the gear speed ratio, gearbox should be operating within a constant operating condition (i.e. torque). As wind power speed fluctuates over time, the variance in wind speed causes the gear to be operating within different operating conditions within relatively short time period (Pengfei & Wang, 2011). As a result, the gear speed ratios may not be a successful indicative of wear, if operated under different conditions (Margaris et al., 2010). However, performing the speed ratio analysis using a consistent operating condition can produce a meaningful measure of gear wear monitoring (Koutras et al., 2007).

In order to analyze the gear ratios, the overall operating time has been divided into modes with similar physical characteristics. The process of splitting data into the operating modes is performed by clustering algorithms, which can be categorized under unsupervised machine learning (Yin et al., 2014).

## 3. Operational Wind Farm Data

The data used in this analysis consist of 5 turbines from EDP Renewables Spain, open data center dedicated to wind energy research. The dataset contains 521, thousand data points of five wind turbines, 2MW production capacity each. The data recordings are for the period of beginning 2016 to end of 2017, 10-minute intervals.

The data recordings contain components of wind turbines illustrated in Table 1, total of 85 data elements, including minimum, maximum and standard deviations of 10-minute intervals. In this paper only a small portion of the data elements have been used which were related to rotor to generator speed ratio and the main factors such as wind speed and power produced.

| Grid Integration | Generator |
|---|---|
| Nacelle | Gearbox |
| Tower | Transformer |
| Pitch | Controllers |
| Rotor | Hydraulics |
| Meteorological | Breaks |

*Table 1. Wind turbine components within the dataset*

## 4. Operating modes and Operational Parameters

Industrial equipment and processes run in multiple operational modes, depending on the physical characteristics of process input (e.g., wind speed) and output (e.g., electrical generation) parameters (Pandit & Infield, 2018a). The operating modes may include stop, start, partial or full production, depending on the actual operating requirements such as electrical demand from grid operators. In order to analyze industrial operations, the data points must be grouped based on natural characteristics implied within data values, which uniquely identify the operating mode (Xie et al., 2013). Data points from multiple operating modes, can be combined, for analysis if the operating modes have similar characteristics such as power generation and wind speed (Turnbull et al., 2020), specifically towards the objectives of the analysis.

Using the parameters defined in Figure 1, wind turbine operating modes can be identified using statistical clustering methods. For clustering the wind turbines dataset four parameters illustrated in Figure 1, were selected as the following:

- Grd_Prod_Pwr_Avg: Active power produced and transferred to grid in kilowatts
- Amb_WindSpeed_Avg: Metrological ambient wind speed in meters per second
- Rtr_RPM_Avg: Speed of rotor rotation (revolutions per minutes)
- Blds_PitchAngle_Avg: often shortened to pitch is the angle between the blade chord line and the plane of rotation, measured in gradient

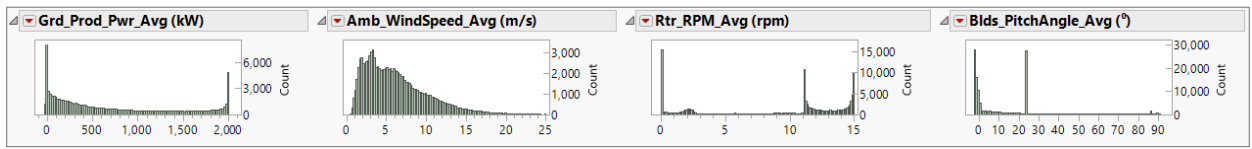

*Figure 1. Histogram of four main wind turbine parameters for identification of operating modes for turbine 1*

### Power Curve Analysis

Power curves (active power generated versus wind speed) have been used within the wind energy industry for performance measuring and production prediction under different wind regimes (Sohoni et al., 2016). Table 2, illustrates the power curves of the wind turbines within the analysis dataset. Although power curves are similar in shape, they also exhibit differences which analysis of the detailed cause, are out of scope for this paper. As illustrated in Table 2, Turbine T06 exhibits the most clearly defined power curve with least deviation from the main operating range. Reasons for the differences are usually within the control system which might have been influenced with turbine operating conditions (e.g., equipment maintenance status) or even differences in wind regime (e.g., wake effects) (Yang et al., 2013).

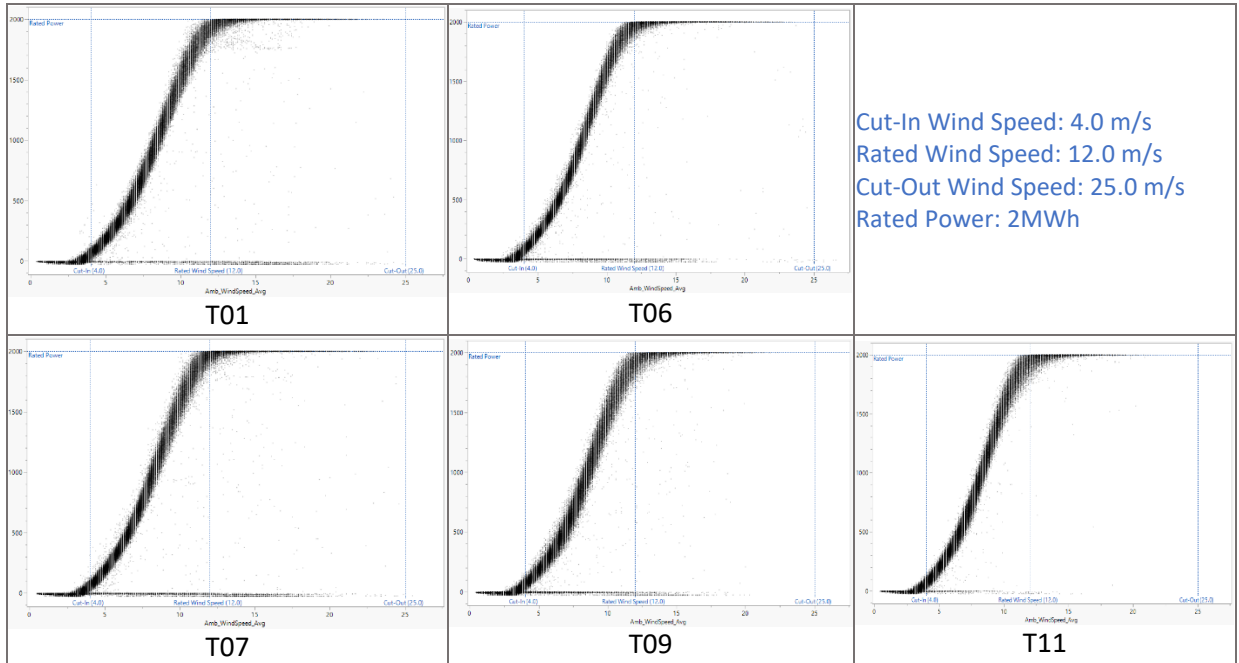

*Table 2. Power Curve of all wind turbines*

Although power curves describe the main input and output factors; however, do not explain the internal operations and control mechanisms. For instance, as wind increases in speed at idling mode, the rotor rotation speed is increased and upon reaching a predefined threshold, generator is connected to rotor to start generating electricity. As wind speed becomes stronger, the pitch control is used for reducing the harvested wind power; ultimately for stronger winds the turbine is taken out of operation for safety and operation integrity. The process of establishing torque transfer from rotor to gearbox and consequently to generator, managing operations using blade angle - although influenced the operations, cannot be seen in the power curve.

Wind energy industry has been reviewing power curve designs for effectiveness in analysis (Sohoni et al., 2016), exploring methods for using new analytics techniques (Yang et al., 2014). For this analysis, a third element of rotor speed has been added as the third axis to the power curve visualization, while four parameters have been used to cluster the overall operational data points (fourth element being blade or pitch angle). Figure 4, illustrates the extended power curve, including the rotor speed, in a three-dimensional scatterplot space. As can be seen the extended power curve includes effects of rotor speed among with generated power and wind speed more clearly.

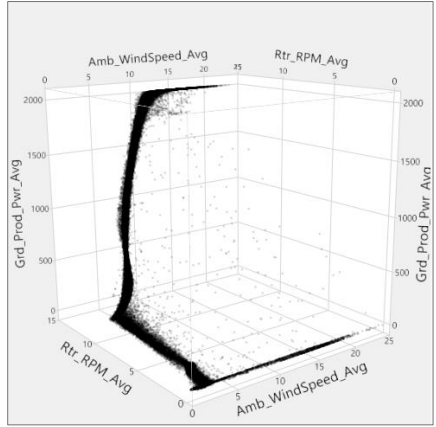

*Figure 2. Extended power curve, to include rotor speed (RPM)*

## Clustering Operational Modes

Clustering is an unsupervised machine learning technique, used for grouping of data points, which have similar characteristics. For wind turbine analytics the similarities are based on the four identified parameters, which are also closely related to the operational modes of the turbines. The algorithm used in this analysis is Normal Mixture, which is an effective classification algorithm for large datasets with overlapping clusters (Fraley & Raftery, 2007).

Normal Mixture uses AIC (Akaike information criterion) and BIC (Bayesian information criterion) as measuring indicators of fit among clustering models (Aho et al., 2014). Both AIC and BIC are known as penalized-likelihoods, sum of errors, to the center of clusters, with different sensitivity towards number of parameters. BIC penalized models with larger parameters, while disadvantageous for analysis of larger data sizes (Brewer et al., 2016). Since the turbine data clustering comprised of only four parameters and large number of data points, both methods of AIC or BIC result very similar measurements of fit (Proust, 2021). Authors selected AIC as indicator of fit for this paper, which produced similar results as BIC. Discussion of selecting the best clustering method and indicator of fit can be considered an independent research topic and have been excluded from this paper.

Figure 3, illustrates AIC versus number of clusters for all turbines. Numerically the most effective number of clusters is the lowest AIC, which is 20 clusters. However, operationally managing 20 clusters might be impractical and not the proper number of operational modes for a wind turbine. Perhaps the right number of operational modes is between 6 to 8 clusters, as the drop in AIC is decreasing with increasing number of clusters. If too many clusters are selected, small operational concentrations due to equipment maintenance conditions might be considered as an operational mode, which is a good example of over-fitting (Vabalas et al., 2019).

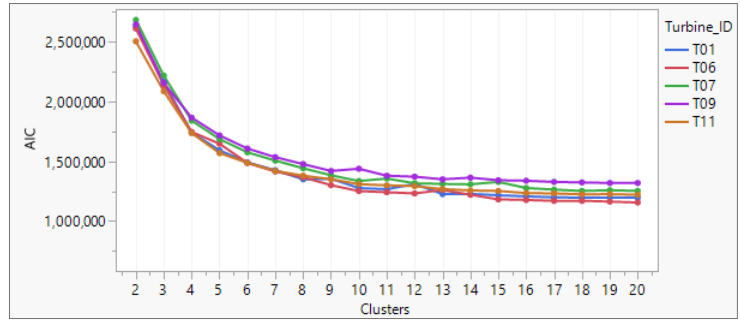

*Figure 3. AIC versus number of clusters*

The objective of clustering is finding the operating modes, for data points with similar characteristics, which will simulate the actual operations over the timeframe of study. For this research six clusters produced a reasonable total error, which each cluster could be considered an operational mode. Interestingly as Figure 3, illustrates, AIC for Turbine T01 is close to lowest and for turbine T09 the highest total error respectively. Lower AIC for turbine T01 means the data points could be assigned to clusters with less error for turbine T01 than turbine T09.

Table 3, illustrates the clusters identification by Normal Mixture algorithm for turbine T01. For this study six clusters as operational modes were selected, which has been illustrated in details (Figure 4). Reviewing the charts, seven or eight clusters very correctly describe the operations as well, which might be operationally more accurate. However, for the purpose of this study six operational modes were accurate enough, since did not need to differentiate between two classes of idling modes. Reviewing the operational diagram with eight clusters although, splits Sub-Rated data points, while might be a correct assessment is not the intention of this study.

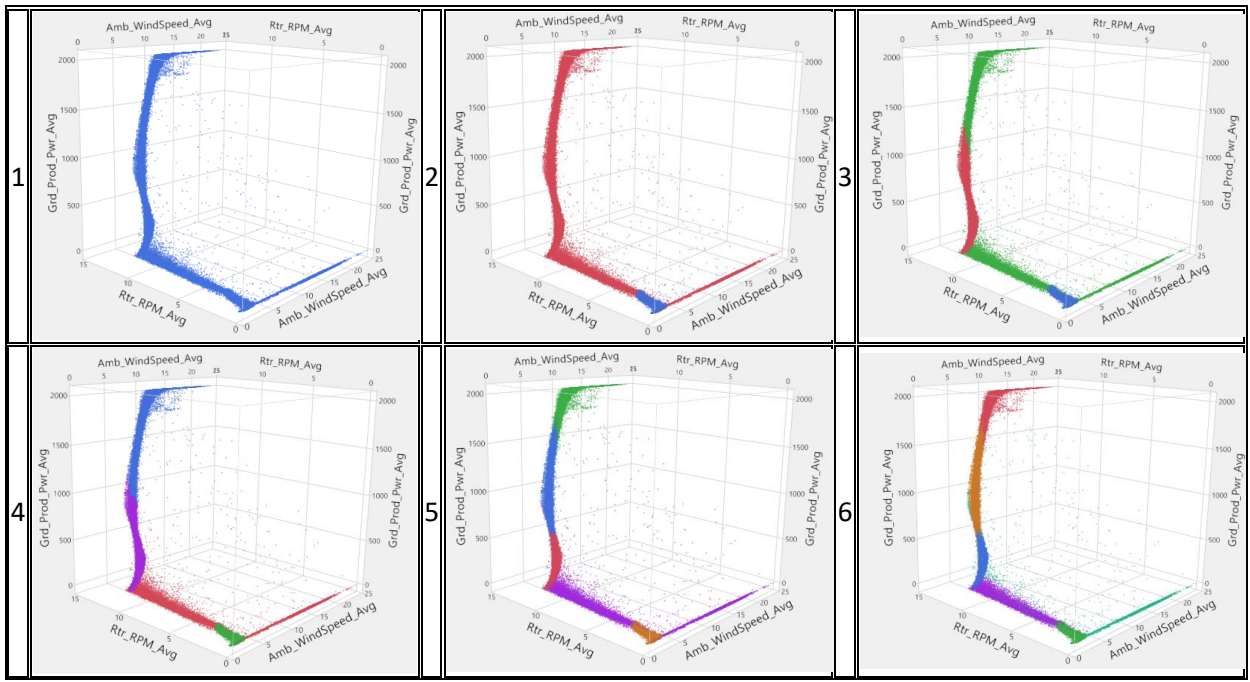

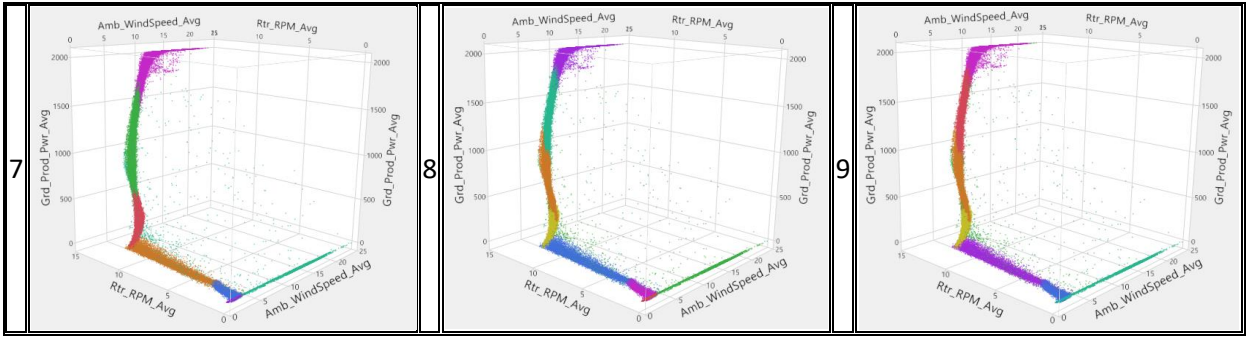

*Table 3. Turbine 1 clusters identified by Normal Mixture algorithm*

Table 4, contains the four main parameters, minimum, maximum and average for each identified operating mode. As indicated, Pitch Managed operation uses higher average blade angle in order to reduce the harvested wind energy which can be used for reducing the rotor speed and ultimately stopping electricity production for stronger wind speeds or going to maintenance mode.

| Operating Mode | Count | Wind Speed (m/s) | | | Rotor Speed (RPM) | | | Blade Angle (Grad) | | | Active Power (kW) | | |
|---|---|---|---|---|---|---|---|---|---|---|---|---|---|
| | | Min | Max | Mean | Min | Max | Mean | Min | Max | Mean | Min | Max | Mean |
| Grid Connecting | 29548 | 3.1 | 6.8 | 5.1 | 10.5 | 13.3 | 11.5 | -2.3 | 0.8 | -1.1 | -0.9 | 579.2 | 223.8 |
| Idling | 26767 | 0.4 | 4.6 | 2.1 | 0 | 4.3 | 0.8 | 23.6 | 24.3 | 23.9 | -25 | 0.9 | -5.7 |
| Sub-Rated Prod | 22993 | 5.6 | 10.6 | 8.1 | 11.8 | 14.9 | 13.9 | -2.5 | -0.4 | -1.9 | 91 | 1710.3 | 923.0 |
| Pitch Managed | 3094 | 0.6 | 24.8 | 8.6 | 0 | 14.9 | 2.5 | -2.2 | 90.6 | 65.6 | -30.1 | 1803.9 | 84.7 |
| Rated Production | 13958 | 9.2 | 23.5 | 12.6 | 14.3 | 14.9 | 14.8 | -2 | 22.8 | 4.1 | 1322.2 | 2000.5 | 1870.1 |
| Start | 8323 | 1.7 | 5 | 3.4 | 0 | 11.5 | 7.1 | -0.4 | 36.4 | 11.0 | -27.5 | 122 | 11.3 |

*Table 4. Physical characteristics of each operating mode*

Figure 4, illustrates the operating mode within the three-dimensional power curve plot, with each mode being separately labeled.

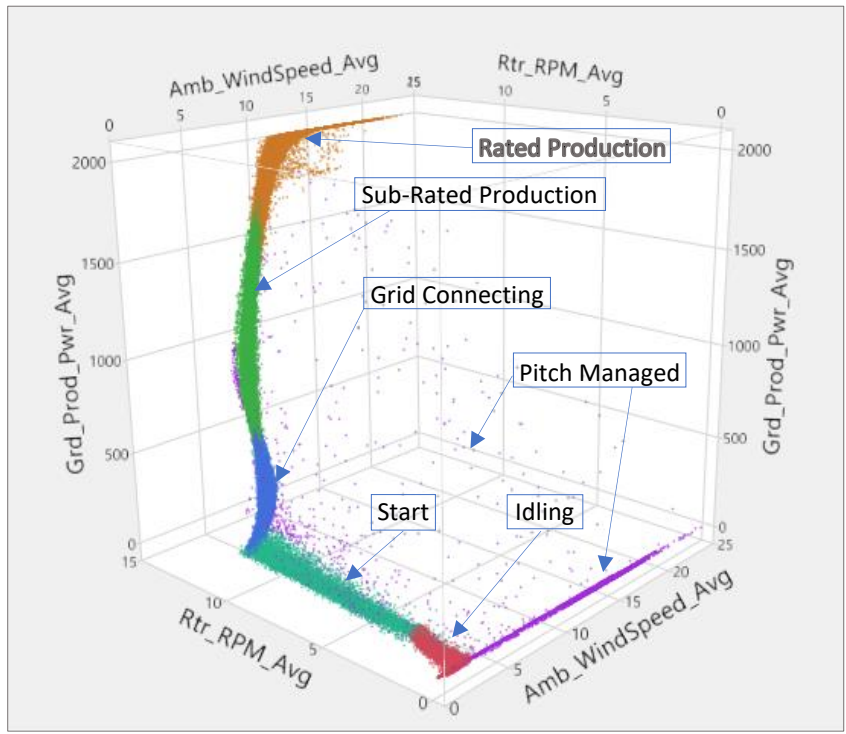

*Figure 4. Extended power curve including rotor speed*

Table 5, illustrates extended power curve of all turbines among with their identified clusters. As can be seen the identified clusters are presenting similar operating modes, with minute differences which are mostly due to differences in operations, or wind regime such as wake effects.

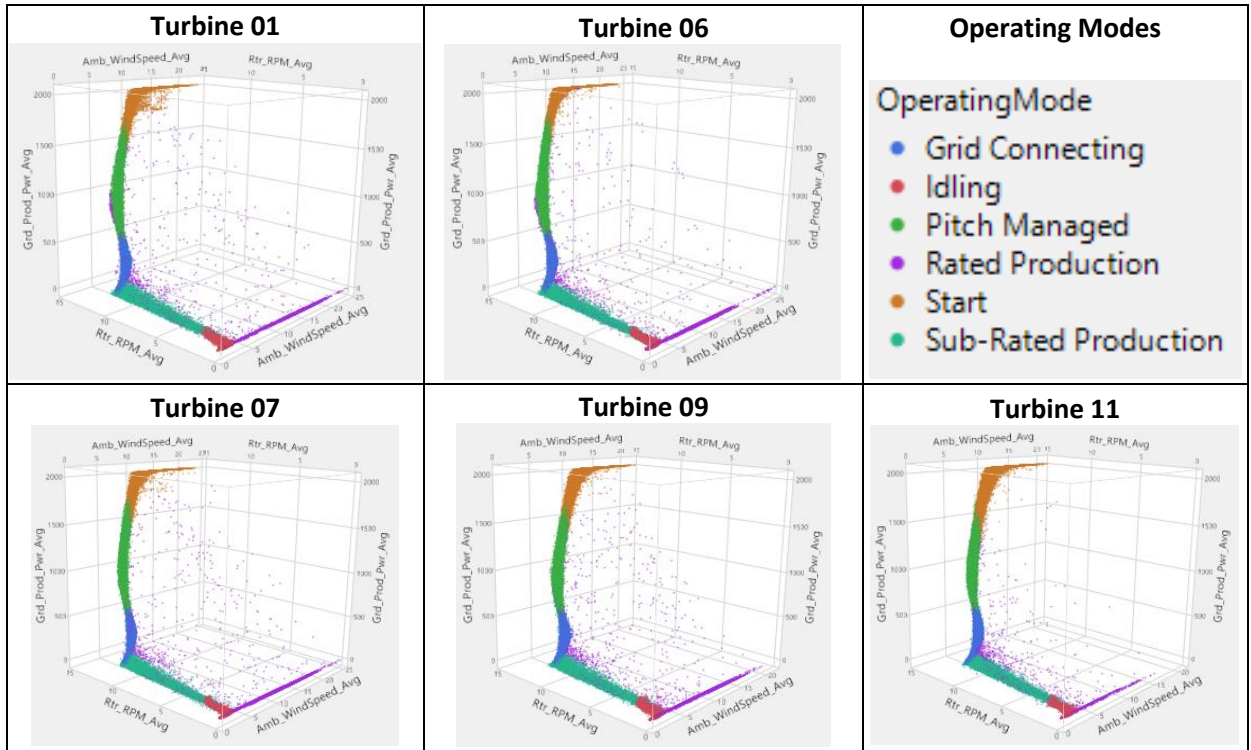

*Table 5. Extended power curve of all turbines including identified clusters*

## Operational Modes and Physical Characteristics

Each different operational mode causes different types of gears behavior based on wind intensity, torque and vibration frequencies (Table 6). The data available for this analysis based on 10 minutes intervals cannot be used for analysis of dynamic forces for each operational mode; however, can be used for measuring gear ratios, as an indication of gear tooth wear over time.

| Operational Mode | Load State |
|---|---|
| Sub-Rated Production | • Wind speed enough for power generation<br>• Pitch control not used to gain maximum power |
| Rated Production | • Maximum wind speed and torque |
| Start | • Increasing/decreasing power production and torque<br>• Extreme forces on gear tooth |
| Idling | • Continuously fluctuating rotational speed and torque<br>• Minimum wind speed and no electrical generation |
| Pitch Managed | • Pitch at high angles to reduce speed or maintaining parking position<br>• Transition between modes for slowing down or parking mode |
| Connecting to Grid | • Increasing rotor speed with minimum torque<br>• Wind speed meets minimum requirement for power generation<br>• Gradually connecting to grid for electrical generation |

*Table 6. Operational modes used in analysis including physical characteristics and potential damages*

# 5. Gearbox Ratio Analysis

Gearbox ratio is defined as number of driving gear revolutions per minute divided by driven gear revolutions (Amarnath et al., 2011). Gearbox ratio, although is very steady over lifetime of equipment, however minute changes in the ratio can be detected, which can be considered as a condition-based monitoring indicator of gear tooth health (Qiu et al., 2017).

There have been extensive research into gear tooth wear, fatigue, pitting and microcracks conducted with reliability engineers for decades. In order to analyses different types of gear tooth failures, more detailed dataset with higher sampling frequency in addition to oil analysis is required (Zeng et al., 2020). Given the available dataset, analysis of gear ratio change over time is feasible, which is an indication of gear tooth wear (Jiang et al., 1999).

As over time wind turbine gearboxes are used under turbulent wind conditions, microcracks change shape of gear teeth, which could be caused by extreme pressures, torque reversals, manufacturing impurities, misalignments or other factors (Z. Zhang et al., 2017). This study is focused towards condition monitoring of gearbox ratio, independent of the cause and remediation techniques.

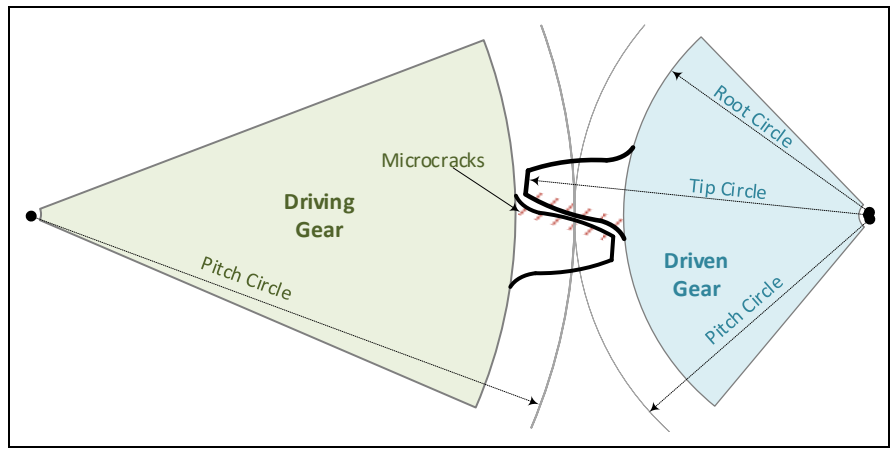

Figure 5. Gearbox tooth contacts and ratio measurements

As wear and fatigue reshapes the gear tooth, pitch diameter changes, due to gradual shift in the contact point between the two gear teeth (Ranganai, 2016). As the contact point between two gears shifts over time, hence the speed ratio also changes, which can be measured and monitored under appropriate conditions, using the rotation speed of the gears (Zaher et al., 2009). Formula 1, shows the relationship between gear speed and radius, hence as gear radius change due to wear over time, the speed also effectively changes.

$$Speed\ Ratio = \frac{Rotation\ Speed\ of\ Driving\ Gear\ (RPM)}{Rotation\ Speed\ of\ Driven\ Gear\ (RPM)} = \frac{Driving\ Pitch\ Radius}{Driven\ Pitch\ Radius}$$

Formula 1. Gearbox speed ratio and radius

Table 7, illustrates graphical plot of rotor speed versus gear speed for different operating modes. As the plots illustrate, two modes of idling and parking do not provide any relatively accurate estimation for the speed ratio. Reviewing $R^2$ also indicated the selected operational modes for the analysis maintain 99.6% accuracy. Since condition based monitoring attempts to detect small variations from normal,

therefore, in author's opinion minimum R² for building models should be consideres as 99% accuracy – although this claim requires further quntification based on actual avidence. As a result, the two operating modes of Idling and Rated Production were removed from more detailed analysis.

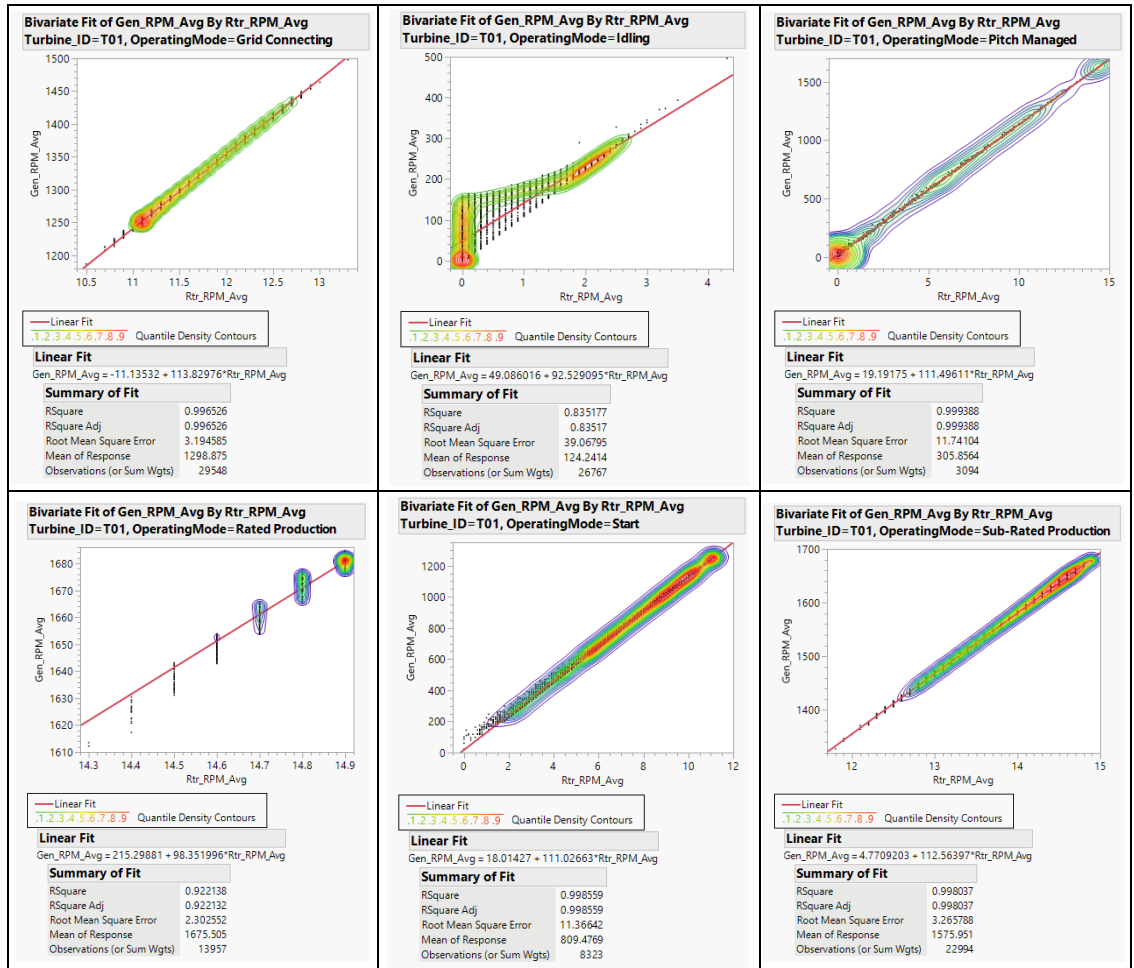

Table 7. Rotor speed versus gear speed rotation plot

# 6. Gear Ratio Change Analysis

## Analysis of Speed Ratio Residuals

Figure 6, illustrates the residual (actual – predicted) of gear speed ratio over time. The first year of 2016, was used as training of simple linear regression model and the last year (2017) was used as validation. Linear regression mathematical models, result residuals with average zero over the entire training period (Flodin & Andersson, 1997). Meaning, the first year (2016) total sum of residuals are zero. The second year (2017) is not restricted to have total sum error of zero.

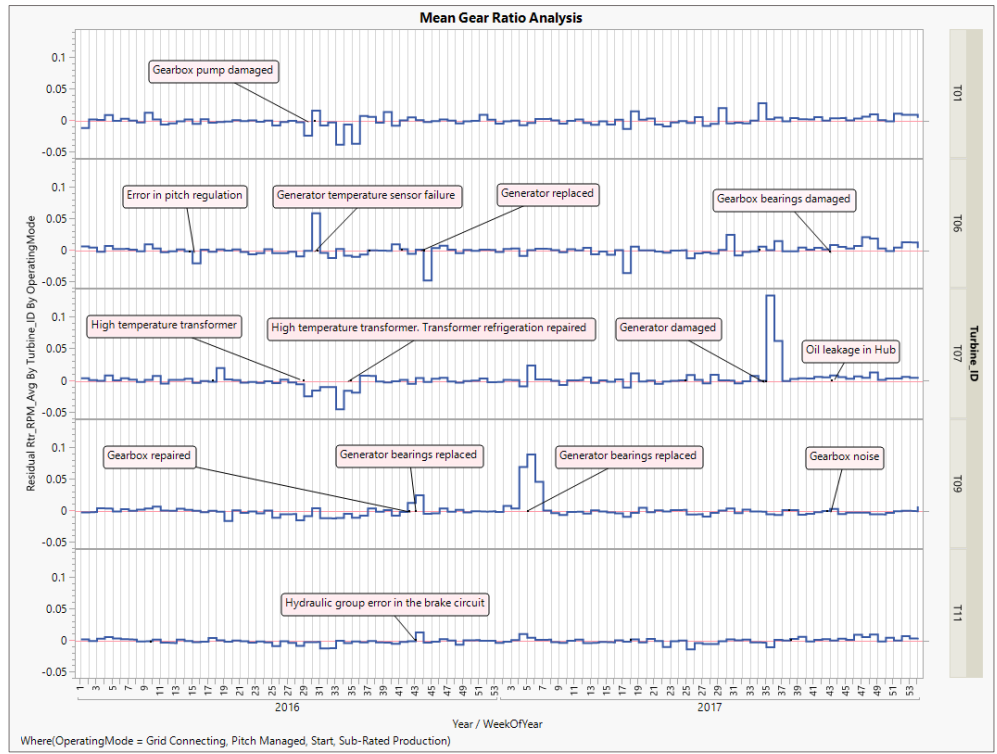

*Figure 6. Mean residual of gear ratio (actual - predicted)*

As illustrated in Figure 6, gear ratio analysis highlights variety of failures, which may happen in the drive-train. For instance turbine T09, year 2016, week 42 is a clear indication of gearbox issues and increase in mean residuals. Same turbine T09, year 2017, week 43, also shows a gradual increase in mean residuals, prior to detection of gearbox noise. Turbine T11, did not have any major issue related to the gearboxes, while other turbines encountered continuous generator failures which can be in Figure 6.

# 7. Conclusion

Condition based monitoring requires precise and accurate measurement data captured over a time period covering variety of conditions (e.g., wind speed and power production), as well as failures in order to be able to detect minute change in the behavior of overall process.

Each operating modes highlights different capability of a process (i.e. wind turbine). Therefore, in order to monitor a specific characteristic of the equipment, similar operating condition should exist. This study used operating modes, by means of clustering to identify and isolate similar conditions. Monitoring gear speed ratio under similar pressure conditions, in fact provides an indication of change in the contact point of gear tooth due to wear.

The main result from this study is dividing the overall operations into smaller section modes by means of clustering, then removing the clusters with lower $R^2$, predicting the outcome. Author used 99% accuracy threshold predicting outcome, in order to keep the cluster within analysis.

## 8. Future Analysis

The analysis included in this paper is based on monitoring gear speed ratio for specific operating modes and conditions. Other condition-based monitoring also could be used such as monitoring performance of electricity production for specific operating conditions such as wind speed and ambient temperature. The challenge with performance-based monitoring is modelling a complex process considering the cooling system, hydraulics, break mechanism, set-points and internal control systems which may affect the performance of electrical generation, which might not be very successful, considering the high $R^2$ needed for condition based monitoring.

Control charts offer graphical visualization of process change over time, by using variety of statistical methods. Control charts can be used for instance monitoring wind turbine performance as well as gear ratio change over time. The advantage of using control charts is the capability of monitoring process behavior based on statistical requirements defined by a set or parameters such as upper or lower limits, standard deviations and statistical distributions (Koutras et al., 2007).

## 9. Acknowledgements

## 10. References

Aho, K., Derryberry, D., & Peterson, T. (2014). Model selection for ecologists: the worldviews of AIC and BIC. Ecology, 95(3), 631–636. https://doi.org/10.1890/13-1452.1

Amarnath, M., Chandramohan, S., & Seetharaman, S. (2011). Experimental investigations of surface wear assessment of spur gear teeth. Journal of Vibration and Control, 18(7), 1009–1024. https://doi.org/10.1177/1077546311399947

Brewer, M. J., Butler, A., & Cooksley, S. L. (2016). The relative performance of AIC, AICC and BIC in the presence of unobserved heterogeneity. Methods in Ecology and Evolution, 7(6), 679–692. https://doi.org/10.1111/2041-210x.12541

Du, M., Yi, J., Mazidi, P., Cheng, L., & Guo, J. (2017). A Parameter Selection Method for Wind Turbine Health Management through SCADA Data. Energies, 10(2), 253. https://doi.org/10.3390/en10020253

Duda, T., Jacobs, G., & Bosse, D. (2018). Investigation of dynamic drivetrain behaviour of a wind turbine during a power converter fault. Journal of Physics: Conference Series, 1037, 052031. https://doi.org/10.1088/1742-6596/1037/5/052031

Flodin, A., & Andersson, S. (1997). Simulation of mild wear in spur gears. Wear, 207(1–2), 16–23. https://doi.org/10.1016/s0043-1648(96)07467-4

Fraley, C., & Raftery, A. E. (2007). Bayesian Regularization for Normal Mixture Estimation and Model-Based Clustering. Journal of Classification, 24(2), 155–181. https://doi.org/10.1007/s00357-007-0004-5

Jiang, J., Chen, J., & Qu, L. (1999). THE APPLICATION OF CORRELATION DIMENSION IN GEARBOX CONDITION MONITORING. Journal of Sound and Vibration, 223(4), 529–541. https://doi.org/10.1006/jsvi.1998.2161

Kim, K., Parthasarathy, G., Uluyol, O., Foslien, W., Sheng, S., & Fleming, P. (2011). Use of SCADA Data for Failure Detection in Wind Turbines. ASME 2011 5th International Conference on Energy Sustainability, Parts A, B, and C. Published. https://doi.org/10.1115/es2011-54243

Koutras, M. V., Bersimis, S., & Maravelakis, P. E. (2007). Statistical Process Control using Shewhart Control Charts with Supplementary Runs Rules. Methodology and Computing in Applied Probability, 9(2), 207–224. https://doi.org/10.1007/s11009-007-9016-8

Liu, X., Lu, S., Ren, Y., & Wu, Z. (2020). Wind Turbine Anomaly Detection Based on SCADA Data Mining. Electronics, 9(5), 751. https://doi.org/10.3390/electronics9050751

Margaris, I. D., Hansen, A. D., Sørensen, P., & Hatziargyriou, N. D. (2010). Illustration of Modern Wind Turbine Ancillary Services. Energies, 3(6), 1290–1302. https://doi.org/10.3390/en3061290

Pandit, R. K., & Infield, D. (2018). SCADA-based wind turbine anomaly detection using Gaussian process models for wind turbine condition monitoring purposes. IET Renewable Power Generation, 12(11), 1249–1255. https://doi.org/10.1049/iet-rpg.2018.0156

Pengfei, L., & Wang, W. (2011). Principle, structure and application of advanced hydrodynamic converted variable speed planetary gear (Vorecon and Windrive) for industrial drive and wind power transmission. Proceedings of 2011 International Conference on Fluid Power and Mechatronics, 839–843. https://doi.org/10.1109/fpm.2011.6045878

Proust, M. (2021). Predictive and Specialized Modeling (16.0). JMP, A Business Unit of SAS.

Qiu, Y., Chen, L., Feng, Y., & Xu, Y. (2017). An Approach of Quantifying Gear Fatigue Life for Wind Turbine Gearboxes Using Supervisory Control and Data Acquisition Data. Energies, 10(8), 1084. https://doi.org/10.3390/en10081084

Ranganai, E. (2016). On studentized residuals in the quantile regression framework. SpringerPlus, 5(1). https://doi.org/10.1186/s40064-016-2898-6

Sohoni, V., Gupta, S. C., & Nema, R. K. (2016). A Critical Review on Wind Turbine Power Curve Modelling Techniques and Their Applications in Wind Based Energy Systems. Journal of Energy, 2016, 1–18. https://doi.org/10.1155/2016/8519785

Turnbull, A., Carroll, J., & McDonald, A. (2020). Combining SCADA and vibration data into a single anomaly detection model to predict wind turbine component failure. Wind Energy, 24(3), 197–211. https://doi.org/10.1002/we.2567

Vabalas, A., Gowen, E., Poliakoff, E., & Casson, A. J. (2019). Machine learning algorithm validation with a limited sample size. PLOS ONE, 14(11), e0224365. https://doi.org/10.1371/journal.pone.0224365

Wang, K., Riziotis, V. A., & Voutsinas, S. G. (2016). Aeroelastic Stability of Idling Wind Turbines. Journal of Physics: Conference Series, 753, 042008. https://doi.org/10.1088/1742-6596/753/4/042008

Wilkinson, M., Darnell, B., Delft, T., & Harman, K. (2014). Comparison of methods for wind turbine condition monitoring with SCADA data. IET Renewable Power Generation, 8(4), 390–397. https://doi.org/10.1049/iet-rpg.2013.0318

Xie, L., Lin, X., & Zeng, J. (2013). Shrinking Principal Component Analysis for Enhanced Process Monitoring and Fault Isolation. Industrial & Engineering Chemistry Research, 52(49), 17475–17486. https://doi.org/10.1021/ie401030t

Yang, W., Court, R., & Jiang, J. (2013). Wind turbine condition monitoring by the approach of SCADA data analysis. Renewable Energy, 53, 365–376. https://doi.org/10.1016/j.renene.2012.11.030

Yin, S., Ding, S. X., Xie, X., & Luo, H. (2014). A Review on Basic Data-Driven Approaches for Industrial Process Monitoring. IEEE Transactions on Industrial Electronics, 61(11), 6418–6428. https://doi.org/10.1109/tie.2014.2301773

Zaher, A., McArthur, S., Infield, D., & Patel, Y. (2009). Online wind turbine fault detection through automated SCADA data analysis. Wind Energy, 12(6), 574–593. https://doi.org/10.1002/we.319

Zeng, X., Yang, M., & Bo, Y. (2020). Gearbox oil temperature anomaly detection for wind turbine based on sparse Bayesian probability estimation. International Journal of Electrical Power & Energy Systems, 123, 106233. https://doi.org/10.1016/j.ijepes.2020.106233

Zhang, Z. Y., & Wang, K. S. (2014). Wind turbine fault detection based on SCADA data analysis using ANN. Advances in Manufacturing, 2(1), 70–78. https://doi.org/10.1007/s40436-014-0061-6

Zhang, Z., Zheng, C., Wen, M., Yang, S., Li, H., & Du, Q. (2017). A Wear Prediction Model for Spur Gears Based on the Dynamic Meshing Force and Tooth Profile Reconstruction. Proceedings of the 2nd International Conference on Computer Engineering, Information Science & Application Technology (ICCIA 2017). Published. https://doi.org/10.2991/iccia-17.2017.150